\documentclass[11pt]{article}

\PassOptionsToPackage{hidelinks}{hyperref}
\usepackage[margin=1in]{geometry}
\usepackage[T1]{fontenc}
\usepackage{amsmath,amssymb}
\usepackage{graphicx}
\graphicspath{{figures/}}
\usepackage{booktabs}
\usepackage[round,authoryear]{natbib}
\usepackage{orcidlink}
\usepackage[expansion=false]{microtype}
\usepackage{hyperref}

\newif\ifdraft \draftfalse
\newif\ifleukos \leukosfalse 
\newcommand{\stver}{\ifdraft{\footnotesize\textsf{~[verified]}}\fi}
\newcommand{\stdes}{\ifdraft{\footnotesize\textsf{~[design]}}\fi}

\newcommand{\Yp}{\mathbf{Y}_p}
\newcommand{\Ypp}{\mathbf{Y}'_p}

\title{When is a closed-form RGB$\rightarrow$S/P ratio adequate?\\
A hyperspectral characterization on natural scenes\\ for mesopic display%
\ifleukos\else\thanks{Licensed under CC BY 4.0.}\fi}

\ifleukos
\author{}
\else
\author{Naoyuki Uchida\\
  \small Independent Researcher, Japan\\
  \small \orcidlink{0000-0002-9551-0470}~\href{https://orcid.org/0000-0002-9551-0470}{0000-0002-9551-0470}\quad
  \href{mailto:naoyuki.uchida.DNF@gmail.com}{naoyuki.uchida.DNF@gmail.com}}
\fi

\date{}

\begin{document}
\maketitle

\begin{abstract}
Mesopic and low-light display transforms require, as their driving signal, a per-pixel
scotopic-to-photopic luminance ratio (S/P); the exact spectral S/P is unavailable for ordinary RGB
content, so a low-cost closed form that estimates S/P from a linear-RGB triplet is used in its place.
Such closed forms exist but have been characterized only on narrowband / LED sources, i.e.\
spectrally sparse spectra, where a relative error of $\approx$41\% has been reported for a
three-channel projection. Display content, however, is natural and broadband. We ask whether the same closed form is \emph{adequate}
there, using per-pixel spectral S/P from hyperspectral imagery as ground truth. On a daylight
radiance time-series, a six-scalar closed form (three photopic and three scotopic channel weights)
reproduces spectral S/P with a median error of
$\approx$0.07 that is time-invariant once the RGB input is chromatically adapted to D65; evaluated in
un-adapted sRGB the error instead carries a color-temperature tilt across illuminants ($\approx$0.19),
so adaptation is the enabling step for this use case. The result generalizes to an independent fifty-scene set (pooled median
0.024; 45/50 scenes within a pre-registered 0.10 band), with the few exceedances concentrated in
saturated, spectrally-peaky surfaces that approach the narrowband regime (floral close-ups in this
set). The scotopic
weight vector is shown to be primary-model dependent, but the value used here is corroborated by a
primary-free XYZ projection, and the median error stays within the band across all principled
coefficient choices. We do not claim observer-validated appearance
fidelity or adequacy on narrowband sources; both are out of scope. Both outcomes follow from the
same three-channel projection: it is overwhelmed by spectrally sparse inputs and adequate on
spectrally smooth ones.
\end{abstract}


\paragraph{Scope of the claim.}
This note characterizes the \emph{adequacy of a closed-form S/P ratio on natural broadband scenes
after chromatic adaptation}. It does \textbf{not} claim a novel appearance transform,
observer-validated appearance fidelity, or adequacy on narrowband / LED sources --- all explicitly
out of scope.

\section{Introduction}

Mesopic and low-light appearance transforms for displays require, as their driving signal, a
per-pixel \textbf{scotopic-to-photopic luminance ratio} (S/P). The S/P ratio sets how far each
pixel sits along the rod--cone transition and therefore how strongly a transform should desaturate
it, shift it toward the Purkinje blue, and lower its apparent brightness.

The exact, spectral S/P requires the per-pixel spectral power distribution (SPD), which is
unavailable for ordinary RGB content. The practical substitute is a \textbf{closed form} that
recovers S/P from a linear-RGB triplet using a small fixed set of scalars --- three photopic and
three scotopic channel weights. Closed-form scotopic estimation is established prior art, though
not originally in that form. \citet{larson1997} introduced a nonlinear scotopic-luminance
approximation from XYZ for tone reproduction and gave no RGB counterpart, noting that graphics had
no standardized RGB primaries at the time. Standardization on sRGB removed that obstacle, and
\citet{maksimainen2019} subsequently reported a per-pixel S/P from sRGB for mesopic imaging. The
six-scalar linear form is what this note evaluates.

These closed forms, however, have been characterized only on \textbf{narrowband and LED sources},
i.e.\ spectrally sparse inputs. \citet{maksimainen2019} report roughly 41\% relative error on
narrowband LEDs. Because three broad channels cannot resolve a narrow spectral spike, this error is
expected from a low-dimensional projection, and narrowband inputs are among the hardest cases for
any 3-channel projection; the literature documents the failure there.

Display content, by contrast, is \textbf{natural and broadband}: daylight-illuminated
reflectances, smooth radiance spectra, low to moderate chroma. Whether the same low-cost closed form
is \emph{adequate} on this content --- the regime opposite to the documented failure --- and under
what preprocessing, has not been characterized against spectral ground truth across realistic
conditions. This note addresses that gap.

\paragraph{Contributions.}
\begin{itemize}
  \item \textbf{(C1) Closed-form S/P adequacy characterization, with generalization.} On D65-adapted natural
  hyperspectral scenes, a 6-scalar closed-form S/P reproduces the true spectral S/P with a
  \textbf{median error of $\approx$0.07, time-invariant across a daylight time-series}, and
  \textbf{generalizes to an independent 50-scene set} (pooled median $\approx$0.024; 45/50 scenes
  within the pre-registered band). This is the opposite regime from the narrowband failure: where
  the projection is benign, the closed form is accurate.\stver
  \item \textbf{(C2) Chromatic adaptation is the enabling step.} Evaluated in \emph{un-adapted}
  sRGB, the closed form carries a color-temperature-dependent tilt (daytime error $\approx$0.19); a
  standard D65 chromatic adaptation removes that tilt and brings the error to the time-invariant
  $\approx$0.07 floor (Fig.~\ref{fig:time}). The closed form must be evaluated in \textbf{adapted}
  sRGB.\stver
  \item \textbf{(C3) Coefficient non-uniqueness, reconciled.} A ``scotopic-from-sRGB'' weight set
  is not unique: it depends on an unstated \textbf{primary spectral model}, and the blue weight in
  particular spans a wide range across principled choices. Our scotopic weights agree with an
  independent, primary-free least-squares projection of $V'(\lambda)$ onto the XYZ basis; the
  Maksimainen weights sit toward the monochromatic-primary end. Neither is an error.\stver
\end{itemize}

\paragraph{Open gap.}
Prior closed forms are characterized only on narrowband / LED sources (the regime where breakdown is expected). The natural-scene adequacy
regime, the chromatic-adaptation requirement, and the primary-model ambiguity in the scotopic
weights are each unaddressed in the literature; this note addresses all three against
hyperspectral ground truth.

\section{Background and related work}

\subsection{Mesopic and scotopic appearance transforms}
A line of work models how scene appearance changes as luminance drops into the mesopic and
scotopic ranges --- loss of chroma, the Purkinje shift toward blue, reduced acuity and brightness.
Representative treatments include \citet{ferwerda1996}, \citet{pattanaik1998,pattanaik2000},
\citet{durand2000}, \citet{thompson2002}, \citet{shin2004}, \citet{kirk2011}, and
\citet{wanat2014}. This mechanism is established prior art and is included here only as the
\emph{motivation} for needing an S/P signal. What this note evaluates is the \emph{quality} (adequacy) of the S/P signal such
transforms consume as input, not the transforms themselves.

\subsection{Closed-form scotopic estimation from tristimulus values}
\citet{larson1997} approximate scotopic luminance from XYZ, using a nonlinear fit to reflectance
samples rather than a linear channel projection; they published no RGB counterpart.
\citet{maksimainen2019} give a per-pixel S/P from sRGB with scotopic channel weights (normalized)
$0.0070 / 0.5190 / 0.4740$ for R/G/B, validated on narrowband LED sources, where the relative error
was reported to reach $\approx$41\%.\footnote{The difference tabulated in that work is the
absolute difference between the two S/P values, expressed as a percentage; normalizing by the
reference S/P instead gives a mean of $\approx$30\%. Either reading is far above the level observed
here on natural broadband scenes.} \citet{maksimainen2019} is therefore the closest prior closed
form of the type evaluated here, and the direct comparison point for \S3. The operative variable is
the spectral sparsity of the per-pixel SPD (illuminant times reflectance): a narrowband illuminant,
as in their case, and a spectrally-peaky reflectance, as in the floral failures here, fall in the
same hard regime, while broadband illumination on smooth reflectances is the opposite, easy one.

\subsection{Chromatic adaptation}
Von Kries--type adaptation, and its modern Bradford / CAT16 variants
\citep{li2017cat16}, transform tristimulus values from a source white to a target white.
Because the closed form's fixed weights implicitly assume a reference (D65) white, the input must be adapted into that domain before the weights apply: \S4 shows that S/P recovered from
\emph{un-adapted} sRGB inherits the scene's color-temperature bias, and that adapting to D65 is
what makes the closed form behave consistently across illumination.

\subsection{Ground-truth hyperspectral imagery}
Per-pixel spectral S/P, the ground truth here, requires hyperspectral imagery (HSI). We use two
natural-scene HSI sources from the Foster group: the Levada daylight radiance time-series
\citep{foster2016} (used as the controlled and time-series stages), and the 2022 fifty-scene
outdoor reflectance set \citep{foster2022data,fosterreeves2022} (used as the independent
generalization stage, \S4.4). Both are broadband natural scenes on a common 400--720~nm / 33-band grid. The two sources probe complementary independence axes: Stage 2 varies illumination over time on one measured-radiance scene, while Stage 3 varies scene content across fifty reflectance sets rendered under a fixed synthetic D65.
\subsection{Summary: the open gap}
No prior work characterizes the closed form's adequacy on natural broadband scenes, states the
chromatic-adaptation requirement, or treats the primary-model ambiguity in the scotopic weights;
the present note supplies all three.

\section{The closed-form S/P and its coefficients}

\subsection{Definition (6-scalar)}
Let $\mathbf{c} = (r, g, b)$ be the \textbf{D65-adapted linear-sRGB} triplet at a pixel. With the
photopic sRGB luminance weights $\Yp = (0.2126,\, 0.7152,\, 0.0722)$ and the scotopic projection
$\Ypp = (0.0184,\, 0.5673,\, 0.3498)$, the closed-form ratio is
\[
\mathrm{(S/P)}_{\text{closed}}
= \frac{\big(\mathbf{c}\cdot\Ypp\big)\big/\big(\mathbf{1}\cdot\Ypp\big)}{\mathbf{c}\cdot\Yp},
\]
where $\mathbf{1}=(1,1,1)$ normalizes the scotopic term so that the equal-energy white maps to~1.
The numerator and denominator are each white-normalized, so the ratio is dimensionless and equals
1 for white; the overall scale of $\Ypp$ cancels, and only its \emph{direction} enters. We do not
carry the absolute $1700/683$ scotopic--photopic constant --- the ratio of the maximum luminous
efficacies $K'_m/K_m = 1700/683 \approx 2.49$. \citet{maksimainen2019} retain that constant, so on
their physical-ratio convention white reads $\approx$2.5; the two conventions differ only by the
fixed white-normalization factor and are placed on the same footing for the comparison in \S3.3.
We take $\Yp$ from the sRGB standard. \citet{maksimainen2019} use $0.2162$ for the R weight, whose
three coefficients therefore do not sum to unity; the standard values preserve the white
normalization the ratio above depends on. Only the photopic side is affected --- the scotopic
weights compared in \S3.3 are theirs as published.

The ground-truth spectral S/P, against which the closed form is evaluated, is
\[
\mathrm{(S/P)}_{\text{spectral}}
= \frac{Y'/Y'_w}{Y/Y_w},\qquad
Y = \!\int V(\lambda)\,L(\lambda)\,d\lambda,\quad
Y' = \!\int V'(\lambda)\,L(\lambda)\,d\lambda,
\]
with $L(\lambda)$ the per-pixel radiance, $V$ and $V'$ the photopic and scotopic luminous
efficiency functions, and $Y_w, Y'_w$ the corresponding white references --- again white-normalized
to~1. (Both definitions verified symbolically with a computer-algebra system; the reproduction
script is released with the published version.)

\subsection{Derivation of the scotopic weights}
Each sRGB primary is approximated by a single Gaussian SPD matched in chromaticity to that primary
and scaled to its photopic luminance $Y_p[i]$; the scotopic weight is then
$Y'_p[i] = \int V'(\lambda)\,\mathrm{SPD}_i(\lambda)\,d\lambda$. Reproducing $\Ypp$ from this
construction recovers the embedded values to $\max|\Delta|\approx 5\times10^{-5}$, and a
multi-start Nelder--Mead fit converges to the same weights independent of the initial Gaussian
width.\stver{} (Reproduction in \texttt{oklchsv\_scotopic\_reconcile.py}.)

\subsection{Non-uniqueness, and reconciliation with Maksimainen et al.}
A primary spectral model carries continuous degrees of freedom (center wavelength, width, shape),
so the single-Gaussian primary is one principled choice among them rather than a unique one. We compare four paths, each normalized to sum~1 so that direction can be read off the
green/blue split (Table~\ref{tab:weights}).

\begin{table}[tbp]
\centering
\caption{Scotopic-from-sRGB weight reconciliation across four principled paths.}
\label{tab:weights}
\footnotesize
\setlength{\tabcolsep}{5pt}
\begin{tabular}{@{}lccp{3.4cm}@{}}
\toprule
Path & Raw $\Ypp=(R,G,B)$ & Normalized $(R,G,B)$ & Character / source \\
\midrule
XYZ-fit (primary-free, LSQ)      & $(0.0000^{\dagger}, 0.5444, 0.3600)$ & $(0.000, 0.602, 0.398)$ & no primary assumed \\
Gaussian primary (this work)     & $(0.0184, 0.5673, 0.3498)$           & $(0.020, 0.606, 0.374)$ & chromaticity {+} luminance matched \\
Maksimainen et al. (2019)        & $(0.0070, 0.5190, 0.4740)$           & $(0.007, 0.519, 0.474)$ & closest prior work \\
Monochromatic primaries          & $(0.0064, 0.3580, 0.6211)$           & $(0.006, 0.363, 0.630)$ & spectrally literal extreme \\
\bottomrule
\end{tabular}

\smallskip
{\footnotesize $^{\dagger}$XYZ-fit $R = -0.0610$ before clipping to~0. \emph{Raw} $\Ypp$ is the
weight vector as applied in the closed form (any negative $R$ clipped to~0); the \emph{normalized}
direction divides by the $R{+}G{+}B$ sum. Because the closed form white-normalizes by
$\mathbf{1}\cdot\Ypp$, the effective direction is $\mathrm{norm}_1(\mathrm{clip}(\text{raw}))$.
sRGB primary dominant wavelengths: 611 / 549 / 464~nm (R / G / B).}
\end{table}

Three observations follow. First, the published sets differ in \textbf{direction}, not merely
scale: a $\approx$0.10 redistribution from green to blue separates ours from Maksimainen's, so the
gap is not a normalization artifact. Second, the two primary-free / perceptually anchored paths ---
the least-squares XYZ projection and the Gaussian-primary construction --- \textbf{converge}
(G 0.602/0.606, B 0.398/0.374, agreement within $\approx$0.03), which corroborates $\Ypp$
independently of the Gaussian-primary model (one of the two paths, the XYZ projection, uses no primary model at all); the Maksimainen weights instead sit toward the
monochromatic-primary end (consistent with their validation on narrowband LED sources). Third, \textbf{blue is the most model-sensitive coefficient}: moving the
blue-primary Gaussian center from 430 to 490~nm moves the blue scotopic weight from 0.378 to 0.139.
The discrepancy between the two published sets is therefore attributable to differing (and
unstated) primary spectral models, not to an error in either. The R scotopic weight is the least constrained
in every path --- small, and in the XYZ-fit even slightly negative before clipping --- so, within the
tested variants, no claim in this note rests on the R difference.

Table~\ref{tab:weights} is reproduced by the ancillary script
\texttt{oklchsv\_scotopic\_reconcile.py}. The robustness of the \S4 adequacy result to the choice
among these variants is quantified in \S4.5.

\section{Evaluation}

\subsection{Ground truth and metric\stver}
The ground truth is the per-pixel spectral S/P (\S3.1) computed from each hyperspectral scene: the
photopic luminance uses the CMF $\bar y(\lambda)$, the scotopic uses the CIE 1951 $V'(\lambda)$
(\texttt{colour}~0.4.7, \texttt{SDS\_LEFS\_SCOTOPIC['cie\_1951']}), both integrated against the
per-pixel radiance and white-normalized to~1, with the scene white taken as the scene highlight
(L-percentile 97--99; the maximum itself is avoided because it picks up specular reflections and
saturated pixels, so a robust upper band approximates the diffuse white). The closed-form estimate
is computed from the \textbf{D65-adapted
linear-sRGB} render of the \emph{same} scene (radiance $\rightarrow$ XYZ via the CIE CMF
$\rightarrow$ linear sRGB via the standard D65 matrix; Bradford adaptation from the scene white to
D65). The error metric is the per-pixel absolute difference
$|\,\mathrm{(S/P)}_{\text{closed}} - \mathrm{(S/P)}_{\text{spectral}}\,|$, summarized per scene by
its median (the median is the robust statistic; the 90th percentile is reported alongside but, as
noted in \S4.4, its tail is inflated by near-black pixels where the ratio denominator approaches
zero). All six scalars are applied in the adapted-sRGB domain throughout.

The adequacy band used to read the results --- 0.10 in S/P units --- is \textbf{operational}: it is
a working tolerance for the display-transform use case, not an observer-anchored just-noticeable-difference (JND) threshold.
This limitation is stated in \S5. The band and the per-scene pass rate of \S4.4 were
\textbf{pre-registered}, i.e.\ fixed before the Stage-3 set (\S4.4) was run.

The evaluation proceeds in three stages: \textbf{Stage 1} fixes the metric on controlled
illuminants (\S4.2); \textbf{Stage 2} isolates the chromatic-adaptation effect on a natural daylight
time-series (\S4.3); and \textbf{Stage 3} tests generalization on an independent fifty-scene set
(\S4.4).

\begin{figure}[tbp]
\centering
\includegraphics[width=\linewidth]{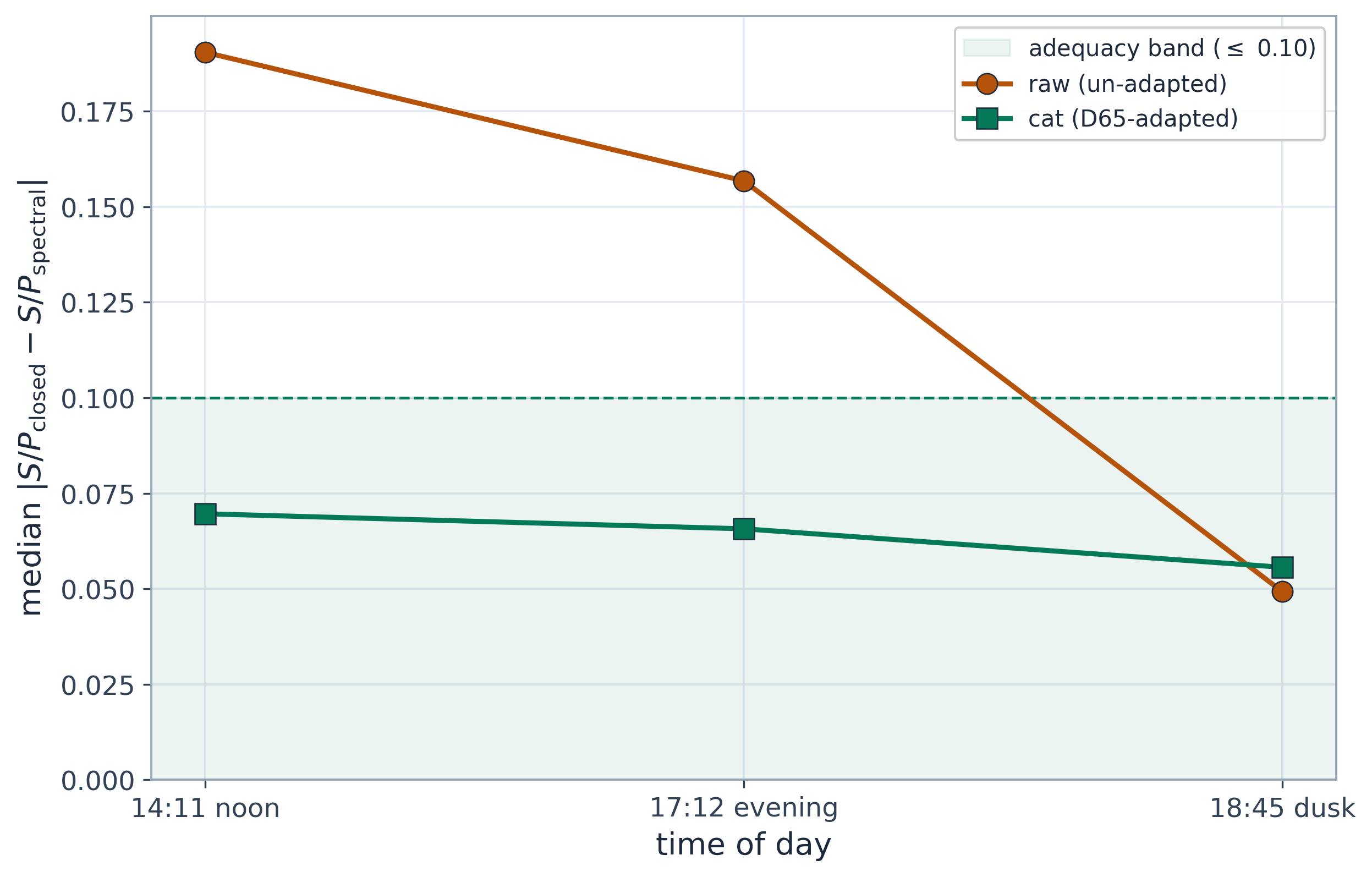}
\caption{Closed-form S/P error versus time of day on the Levada daylight series: un-adapted (sloped,
color-temperature tilt) versus D65-adapted (flat, time-invariant $\approx$0.07 floor).}
\label{fig:time}
\end{figure}

\subsection{Stage 1 --- controlled illuminants\stver}
\begin{figure}[tbp]
\centering
\includegraphics[width=\linewidth]{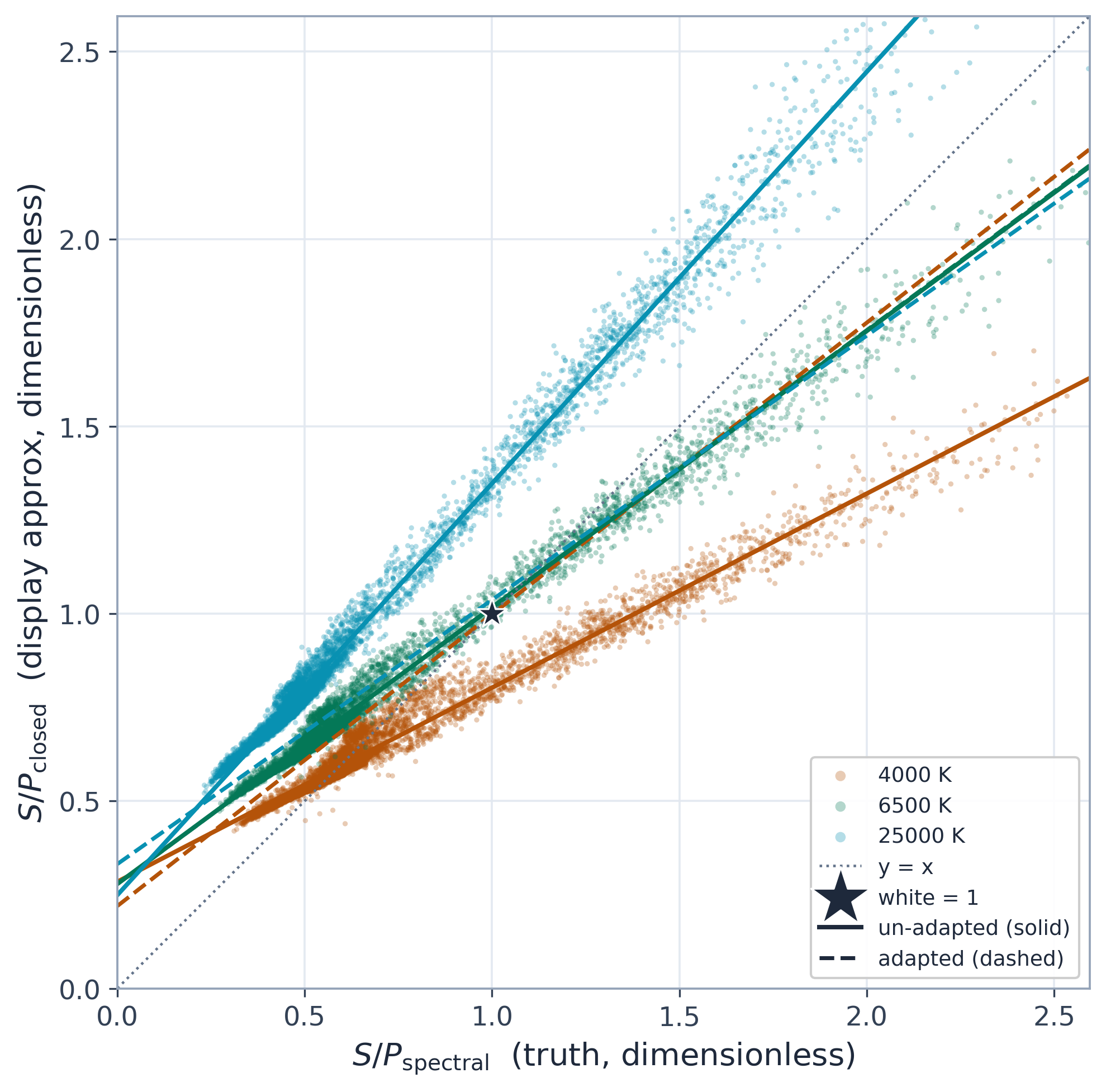}
\caption{Stage 1 --- closed-form vs spectral S/P (three controlled illuminants), with
per-illuminant least-squares fits overlaid (solid: un-adapted; dashed: after Bradford adaptation to
D65). The slope depends on correlated color temperature (steepest at 25000~K, shallowest at 4000~K,
6500~K in between), while the intercept is nearly common across the three illuminants. Adaptation
converges the slopes on $\approx$0.74 rather than unity (Table~\ref{tab:perillum}).}
\label{fig:stage1}
\end{figure}

Stage 1 establishes the metric and the closed-vs-spectral relation on a controlled scene under
known artificial illuminants, before any natural-scene variability is introduced. Within each
illuminant the closed-form estimate is nearly linear in the spectral ground truth (per-illuminant
$R^2 \approx 0.98$), but \textbf{its slope depends on correlated color temperature}: $\approx$1.10 at
25000~K (steepest), $\approx$0.74 at 6500~K, and $\approx$0.52 at 4000~K (shallowest). The
\textbf{intercept, by contrast, is nearly common} across the three illuminants ($\approx$0.25--0.28,
spread $<$0.04); relative to $y=x$ the 25000~K band opens above the identity and the 4000~K band
below it (Fig.~\ref{fig:stage1}; per-illuminant fits in Table~\ref{tab:perillum}). Pooled over the three un-adapted illuminants ($n = 256{,}275$
pixels, no subsampling), the fit gives an overall slope $0.751$, intercept $+0.286$ ($R^2 = 0.735$).
The un-adapted color-temperature dependence thus appears as a \textbf{color-temperature-dependent
slope (gain) about a common intercept}, and this is the structure that \S4.3 shows chromatic
adaptation removes (C2). Stage 1 is therefore a controlled baseline that exposes this
color-temperature-dependent slope and fixes the metric, not an adequacy claim; the adequacy result
proper is carried by the adapted natural scenes of Stages 2--3.

\begin{table}[tbp]
\centering
\small
\caption{Stage 1 per-illuminant regression of closed-form on spectral S/P, before and after Bradford
adaptation to D65 ($n=85{,}425$ pixels per illuminant). Before adaptation the slope depends on
correlated color temperature while the intercept is nearly common; adaptation contracts the slope
spread from $0.581$ to $0.073$ but converges on $\approx$0.74 rather than unity, so the
pre-registered H2 (adaptation restores unit slope) fails.}
\label{tab:perillum}
\begin{tabular}{lcccccc}
\hline
 & \multicolumn{3}{c}{Un-adapted} & \multicolumn{3}{c}{D65-adapted} \\
Illuminant & slope & intercept & $R^2$ & slope & intercept & $R^2$ \\
\hline
4000~K   & 0.518 & 0.285 & 0.975 & 0.778 & 0.220 & 0.985 \\
6500~K   & 0.738 & 0.279 & 0.987 & 0.739 & 0.278 & 0.987 \\
25000~K  & 1.099 & 0.249 & 0.991 & 0.705 & 0.332 & 0.986 \\
\hline
spread   & 0.581 & 0.036 & --- & 0.073 & 0.112 & --- \\
\hline
$|s-1|$  & \multicolumn{3}{c}{0.482 / 0.262 / 0.099} & \multicolumn{3}{c}{0.222 / 0.261 / 0.295} \\
\hline
\end{tabular}
\end{table}

\subsection{Stage 2 --- daylight time-series of a natural scene (Foster Levada)\stver}
Stage 2 evaluates the closed form across time of day on the Levada daylight radiance series
(samples at 14:11 noon, 17:12 evening, 18:45 dusk), where the illumination color temperature drifts
substantially. Two results follow.

First, \textbf{adaptation is the enabling step (C2).} Evaluated in \emph{un-adapted} sRGB, the
closed-form S/P inherits a color-temperature-dependent tilt: the daytime error is $\approx$0.19,
and it varies systematically with the cast (the scene-white correlated color temperature, taken from
the p97--99 highlight, drifts from $\approx$10500~K at noon through $\approx$9400~K to $\approx$5500~K
at dusk, and the un-adapted error tracks it). Applying the standard Bradford adaptation to D65
removes that tilt and brings the error down to the time-invariant floor ($\approx$0.07). The same
adaptation, applied to the controlled illuminants of Stage 1, removes the color-temperature
dependence of the slope: the per-illuminant slope spread contracts from $0.58$ to $0.07$ (about one
eighth; Table~\ref{tab:perillum}). Because the adaptation target D65 corresponds to a correlated color temperature
$\approx$6500~K, the 6500~K illuminant is a near-identity adaptation (slope $0.738 \!\to\! 0.739$),
and the 4000~K and 25000~K slopes converge to this D65-referenced common slope $\approx$0.74. The
residual common slope $0.74$ ($<1$) is color-temperature-independent and reflects a
projection-intrinsic underestimation: even under D65 the three-channel closed form systematically
undershoots the spectral S/P. The pre-registered H2 (adaptation restores unit slope) therefore
\emph{fails}; adaptation removes the color-temperature dependence only.
Fig.~\ref{fig:time} shows the error against time of day for the two cases --- sloped (raw) versus
flat (adapted).

Second, \textbf{after adaptation the error is small and time-invariant (C1, time axis).} The
adapted per-sample median errors are 0.0696, 0.0657, and 0.0556 (noon, evening, dusk); the pooled
median --- the median of the three --- is \textbf{0.0657}. The closed form does not drift as the
daylight color temperature changes, which is the property a display transform requires of its
driving signal.

\subsection{Stage 3 --- generalization to an independent scene set\stver}
Stage 2 characterizes the closed form on one scene across time; Stage 3 tests whether the same
adapted-sRGB protocol holds on scenes the coefficients were never tuned to. We run the identical
pipeline --- band support 400--720~nm / 33 bands, same spectral-S/P definition, same render and
Bradford adaptation, same metric --- on the Foster fifty-scene natural set under a canonical D65
illuminant, importing the Stage-2 functions rather than re-implementing them.

\paragraph{Pre-registered criteria.}
Before running, two pass conditions were fixed: (i) the pooled median error stays within the
operational band ($\le$0.10), and within $+0.03$ of the Stage-2 reference; and (ii) at least 80\%
of scenes have a per-scene median $\le$0.10.

\paragraph{Result.}
The set meets both. The pooled median error is \textbf{0.0245} --- below the Stage-2 reference of
0.0657 (a difference of $-0.0412$), well inside the band --- and \textbf{45 of 50 scenes (90\%)}
fall within 0.10. The mean of the per-scene medians is 0.0397; the pooled 90th percentile is 0.1391
(reported for reference, not a pre-registered criterion). The result is stable to the illuminant choice: re-running under 4000~K and 25000~K
gives pooled medians of 0.0233 and 0.0260 (for reference), so the generalization is not an artifact
of the particular (D65) white chosen. The closed form therefore holds in the regime opposite to the documented
narrowband failure: on independent broadband scenes it stays comfortably within the adequacy band.

Under the canonical D65 illuminant the scene white already sits near D65, so the Bradford step is
near-identity here; Stage 3 thus isolates \textbf{scene generalization of the closed form}, while
the adaptation effect itself remains the Stage-2 result (C2). The lower Stage-3 floor (0.0245 versus
the Stage-2 0.0657) follows from this split rather than contradicting it: Stage 3 tests scene
generalization under clean (near-identity) adaptation, whereas Stage 2 additionally carries the
residual of adapting across a genuine daylight CCT drift. Stage 3 is thus the easier test on
adaptation and the harder one on scene diversity, and the closed form clears both.

\begin{figure}[tbp]
\centering
\includegraphics[width=\linewidth]{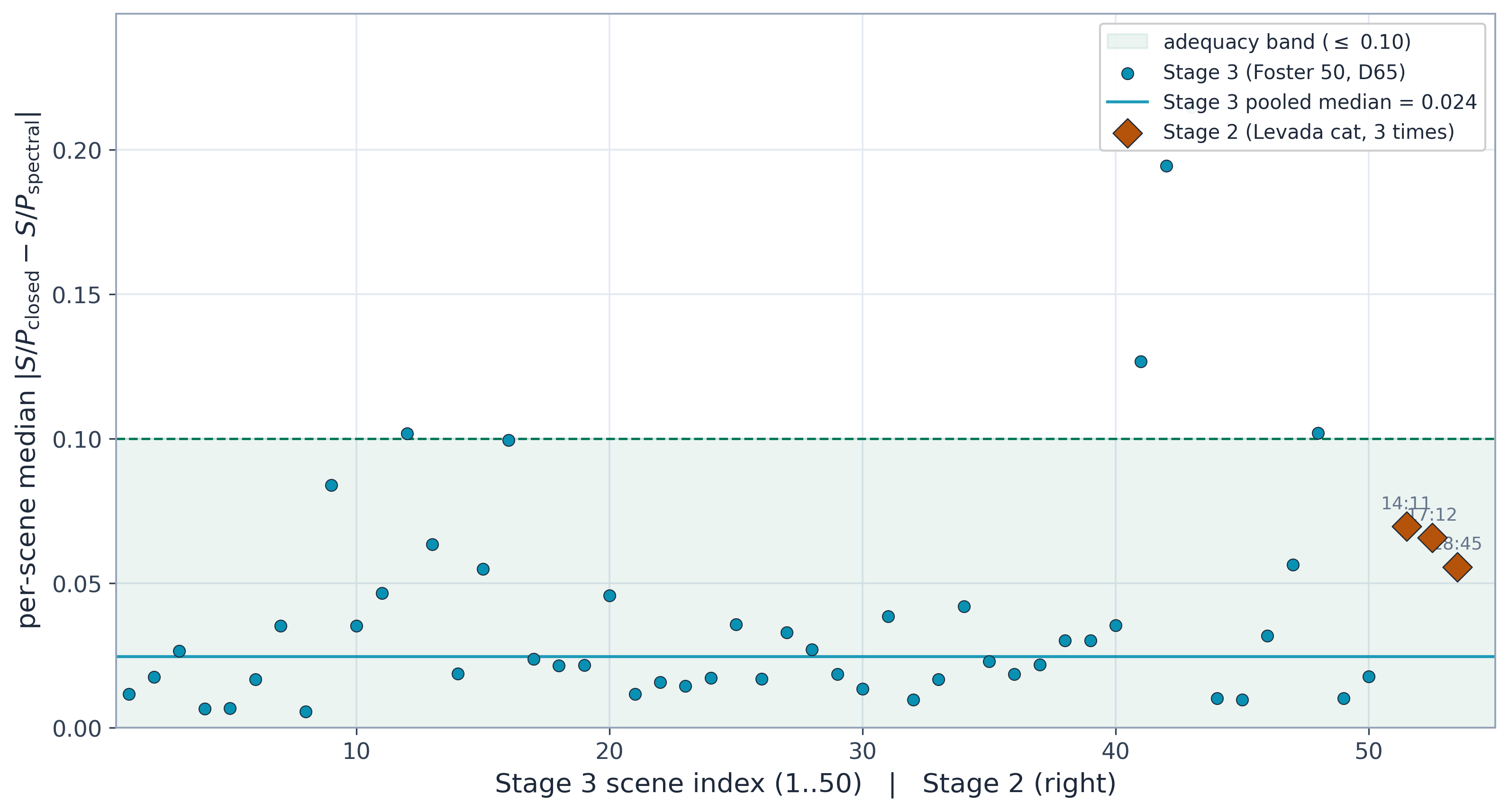}
\caption{Stage 3: per-scene median S/P error across the fifty-scene set against the Stage-2 samples,
with the pre-registered 0.10 adequacy band drawn.}
\label{fig:stage3}
\end{figure}

The five scenes that exceed the band (per-scene medians 0.10--0.21: scenes 43, 42, 41, 48, 12) are
not random failures --- they are high-chroma, spectrally-peaky surfaces approaching the narrowband
regime; their characterization is given in \S5.
Fig.~\ref{fig:stage3} plots the per-scene median error for the fifty scenes against the Stage-2
samples, with the adequacy band drawn.

{\small One metric note for transparency: scene 41's 90th-percentile error is anomalously large
because a cluster of near-black pixels drives the S/P denominator toward zero; its \textbf{median}
(0.1267) is unaffected, which is why the median is the headline statistic and the p90 is reported
only as a tail indicator.}

\subsection{Coefficient-ambiguity robustness\stver}
\begin{figure}[tbp]
\centering
\includegraphics[width=\linewidth]{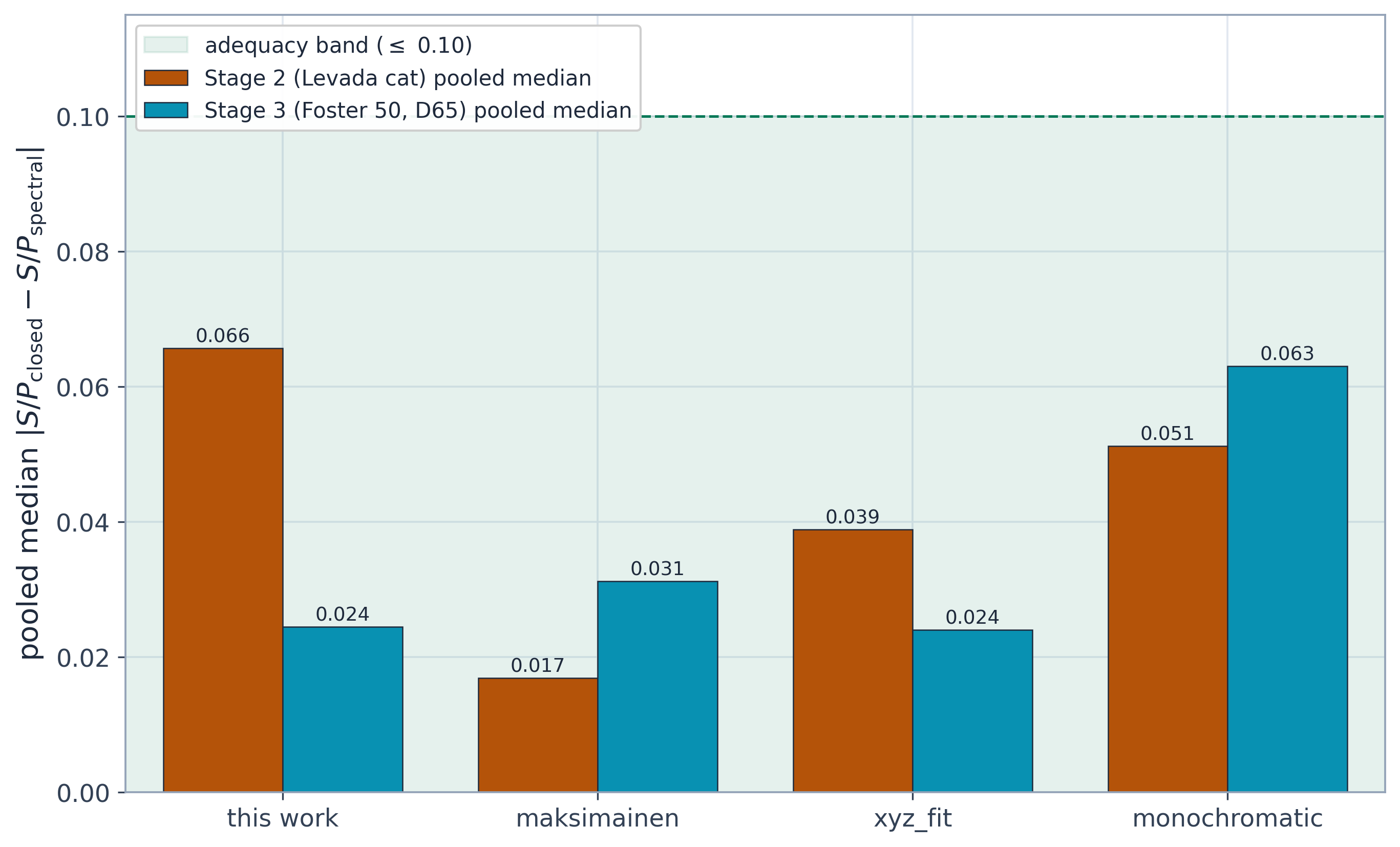}
\caption{Stage 2 and Stage 3 pooled median S/P error under the four principled scotopic-weight
variants of Table~\ref{tab:weights}; all stay within the 0.10 band, the monochromatic extreme
closest to the edge at Stage 3.}
\label{fig:sweep}
\end{figure}

We test how much the \S4 adequacy result depends on the coefficient non-uniqueness of \S3.3 --- the
primary-model dependence of the scotopic weights.
Holding everything fixed except the scotopic weight vector $\Ypp$, we recompute the closed-form S/P
median error under each principled variant of \S3.3 on the identical Stage-2 (cat) and Stage-3
(D65) data; the swept directions are asserted against Table~\ref{tab:weights} before proceeding,
and the baseline reproduces the published pooled medians exactly ($0.0245$, $0.0657$;
$|\Delta|=0$). Pooled medians --- Stage 2: this work 0.066, Maksimainen 0.017, XYZ-fit 0.039,
monochromatic 0.051; Stage 3: this work 0.024, Maksimainen 0.031, XYZ-fit 0.024, monochromatic 0.063.
Every variant stays within the operational band on both stages (across-variant spread 0.049 /
0.039), so the result is bounded within the band, not direction-free. The largest pooled median and
heaviest tail are the spectrally-literal \texttt{monochromatic} extreme (Stage 3 0.063, p90 0.425)
--- the expected extreme. We do not claim our coefficients are optimal: the baseline is in fact the
largest at Stage 2 and not the smallest at Stage 3 (the primary-free XYZ-fit is marginally lower
there). What we claim is \emph{band membership} --- every principled direction stays within 0.10. In
the narrowband regime the projection direction is decisive (a wrong direction breaks the estimate);
on broadband content that direction-sensitivity disappears, and the present result is the broadband
counterpart of that contrast (Fig.~\ref{fig:sweep}).

\subsection{Side observations\stver}
Two secondary effects are recorded for completeness and are not load-bearing for any contribution.
The scene-white luminance $Y_n$ is non-monotone across the Stage-2 series, peaking near golden hour
where direct sun dominates; and the magnitude of the adaptation correction in Stage 2 scales with
the strength of the color cast, consistent with the tilt being a cast-driven effect (\S4.3). Neither
is interpreted further.

\section{Discussion and limitations}

The result of \S4 can be stated concisely: a low-cost closed-form S/P, known to fail on narrowband
sources ($\approx$41\% relative error), is \textbf{adequate on natural broadband scenes} --- median
error $\approx$0.07 on a daylight time-series and a pooled $\approx$0.024 across fifty independent
scenes --- provided it is evaluated in \textbf{D65-adapted} linear sRGB. The scotopic weight that
drives it is primary-model dependent, but the particular value used here is corroborated by an
independent, primary-free projection. Both outcomes follow from the same three-channel projection:
spectrally sparse inputs overwhelm it, spectrally smooth ones do not.\stver

\begin{figure}[tbp]
\centering
\includegraphics[width=\linewidth]{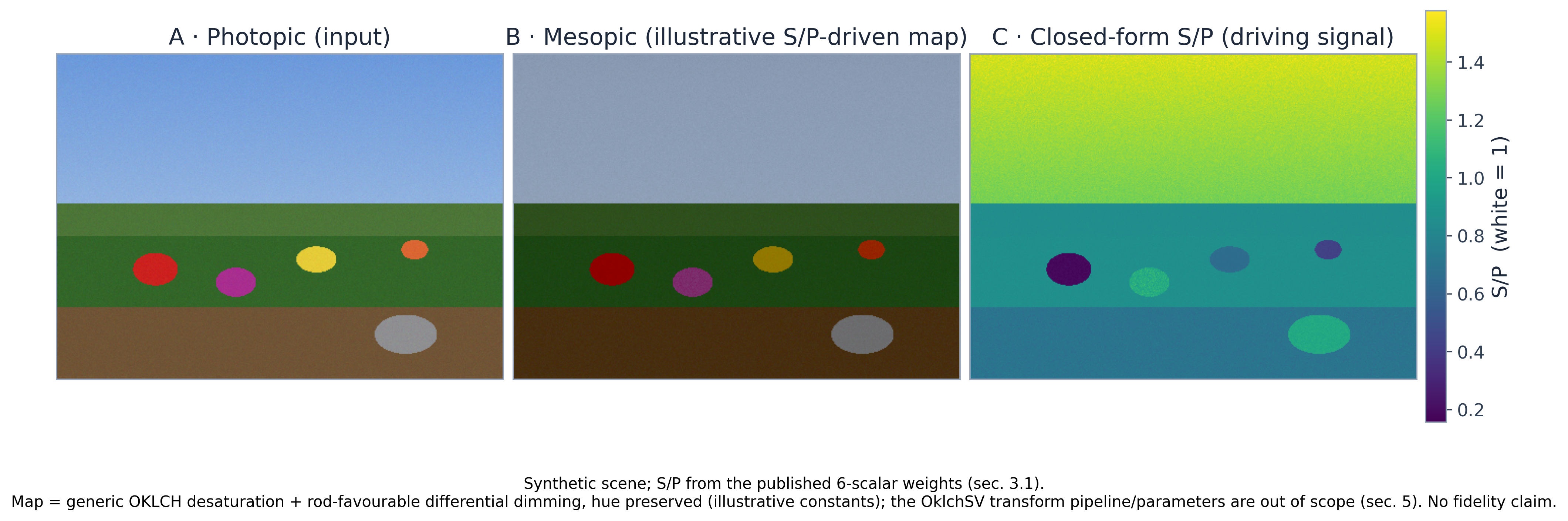}
\caption{Illustrative mesopic appearance change driven by the characterized closed-form S/P.
\textbf{A} photopic input (synthetic broadband scene); \textbf{B} the same scene after a generic,
S/P-driven mesopic map in OKLCH (desaturation and rod-favourable dimming, hue preserved); \textbf{C} the
closed-form S/P field that drives it, from the published 6-scalar weights (\S3.1). OklchSV --- the OKLCH-native mesopic display pipeline this signal feeds --- and its tuned parameters are out of scope (\S5); the mapping constants are illustrative,
not the tuned transform. No fidelity claim is made.}
\label{fig:mesopic}
\end{figure}

\paragraph{Application (illustrative).}
The characterized closed form is what drives an OKLCH-native mesopic display transform; we include
one worked before/after example as an indication of utility (Fig.~\ref{fig:mesopic}). This is an
illustration, not an evaluation: the transform itself is out of scope here, and a full
evaluation, including observer matching, is future work.\stdes

\paragraph{Limitations.}
We state the boundaries of the adequacy claim explicitly.
\begin{itemize}
  \item \textbf{Natural broadband scenes only.} Adequacy degrades on scenes dominated by saturated,
  spectrally-peaky surfaces. Across the fifty-scene set, per-scene error correlates with the 95th
  percentile of Oklab chroma ($r = +0.64$); the five scenes that exceed the 0.10 band are
  high-chroma floral close-ups with narrow, spectrally-peaky reflectance (failing vs.\ passing
  scenes: 95th-pct chroma 0.165 vs.\ 0.082). These pixels approach the narrowband regime, which is
  exactly where a 3-channel projection is known to fail. This is the natural-scene extension of the
  same mechanism, not a separate defect. Narrowband / LED-lit scenes remain out of scope (the
  Maksimainen regime).\stver
  \item \textbf{No observer validation.} The driving signal is evaluated against \emph{physical}
  spectral S/P, not against perceived appearance. The transform it feeds is physiologically
  motivated, not psychophysically matched; appearance fidelity is unverified and is named as future
  work (cf.\ haploscopic matching; \citealp{wanat2014}).\stdes
  \item \textbf{Adaptation dependence.} Adequacy holds only after chromatic adaptation. In
  un-adapted sRGB the closed form inherits a color-temperature tilt (\S4); the $\approx$0.07 floor
  is an adapted-domain result.\stver
  \item \textbf{Coefficient reliability.} The R scotopic weight is the least constrained in every
  derivation path, and the single-Gaussian primary model is itself an approximation. The \S4.5
  sweep bounds how much the adequacy result depends on this; no claim rests on the R weight.\stver
\end{itemize}

The residual common slope $0.74$ ($<1$) that survives adaptation largely reflects a systematic
underestimation: the three-channel projection cannot fully expand the scotopic $V'(\lambda)$ and
photopic $V(\lambda)$ efficiencies in an RGB basis. In principle the loss shrinks if channels are
added in the scotopic-peak region ($\approx$500~nm); since $V'(\lambda)$ is a smooth unimodal curve,
four to five bands covering the blue--green region would already improve it substantially, whereas
extra long-wavelength bands would help little. The catch is that adding channels changes the problem: with multispectral input the closed form is
unnecessary, since S/P integrates directly from the spectrum. So within the actual task --- estimating
S/P from a fixed three-channel display signal --- the $0.74$ loss cannot be removed by more
information; it is a structural floor of that task, consistent with the scope of \S2.1 (the quality
of the S/P signal a transform consumes).
\section{Falsifiability and future validation}

Each contribution is stated so that it can fail against a defined test.
\begin{itemize}
  \item \textbf{C1 (adequacy + generalization)} is refuted if the closed-form S/P error exceeds the
  pre-registered 0.10 operational band --- a working tolerance, not an observer JND --- on the
  Stage-3 set, or on further independent HSI (e.g.\ ICVL,
  \citealp{arad2016}; KAUST-HS, \citealp{li2021kaust}) under the same D65-adapted-sRGB protocol.
  The band exceedance observed here is \textbf{not random}: it is concentrated in a \emph{named,
  predictable} scene class --- high-chroma, spectrally-peaky florals --- consistent with the
  projection mechanism. This yields a testable prediction: saturated-pigment scenes will show larger
  error, broadband scenes will not. A counterexample would be a broadband, low-chroma scene with
  large error, or a saturated, spectrally-peaky scene (florals in this set) with near-zero error.\stver
  \item \textbf{C2 (adaptation is the enabling step)} is refuted if the un-adapted $\rightarrow$
  adapted tilt does not flatten under a standard chromatic adaptation, or if adapting to a
  \emph{non-D65} target flattens it equally well (which would weaken the D65 specificity of the
  claim rather than the adaptation requirement itself).\stver
  \item \textbf{C3 (coefficient reconcile)} claims the two published weight sets arise from
  \emph{different} primary spectral models, and are therefore non-unique rather than in error. It is
  refuted if, to the contrary, a single documented sRGB-primary SPD model can be shown to yield both
  the Maksimainen weights ($0.0070/0.5190/0.4740$) and the least-squares XYZ-fit corroboration of
  $\Ypp$ --- in which case the ``non-uniqueness'' framing would not hold.\stver
  \item \textbf{Robustness.} The median-error result survives the \S4.5 sweep within the stated
  margin: every principled variant stays $\le 0.10$ pooled median on both stages (spread 0.049 /
  0.039). The closest approach to the band edge is the spectrally-literal monochromatic extreme at
  Stage 3 ($0.063$), leaving a margin of $\approx$0.037. The claim fails if any principled variant
  exceeds the band; none did.\stver
  \item \textbf{Recovery by a white-anchored transform (future test).} The residual common slope $0.74$
  is within the operational band but is a systematic undershoot; recovering unit slope would turn
  adequacy into accuracy at no additional free parameter, for uses needing tighter S/P fidelity. It
  can be corrected to unity by a one-parameter transform that fixes the white point
  (S/P $=1$): $y' = 1 + (y-1)/0.74$. Because S/P is normalized to $1$ at white by definition (\S3.1),
  this anchored transform is not a new free parameter but preserves that normalization after
  adaptation, and unlike a proportional scaling it does not worsen the intercept. The correction
  factor $0.74$ was obtained under D65 adaptation on this scene set, however; whether the same factor
  holds across independent scene distributions (generalization) is untested. It is refuted if, on
  independent HSI, the residual after the white-anchored transform leaves the adequacy band, or if the
  common slope departs significantly from $0.74$.
\end{itemize}

In sum, a low-cost closed-form S/P is adequate on D65-adapted natural broadband scenes, fails
predictably on narrowband and saturated-floral inputs, and rests on a primary-model-dependent but
independently corroborated coefficient --- a characterization of an input signal, not an appearance
claim. The open work is an observer-validated evaluation of the transform it feeds and application
to broader independent HSI (ICVL, KAUST-HS), against which the pre-registered band and the named
failure class keep the result falsifiable.


\vfill
\ifleukos\else\noindent\footnotesize Licensed under CC BY 4.0\fi
(\url{https://creativecommons.org/licenses/by/4.0/}).

\end{document}